\documentclass[12pt]{iopart}

\usepackage{amssymb}
\usepackage{psfrag}
\usepackage{graphicx}

\begin{document}

\title{Fundamental properties and applications of quasi-local black
  hole horizons} 
\author{Badri Krishnan} 
\address{Max Planck Institut
  f\"ur Gravitationsphysik, Am M\"uhlenberg 1, D-14476 Golm, Germany}
\ead{badri.krishnan@aei.mpg.de}

\begin{abstract}
  The traditional description of black holes in terms of event
  horizons is inadequate for many physical applications, especially
  when studying black holes in non-stationary spacetimes. In these
  cases, it is often more useful to use the quasi-local notions of
  trapped and marginally trapped surfaces, which lead naturally to the
  framework of trapping, isolated, and dynamical horizons. This
  framework allows us to analyze diverse facets of black holes in a
  unified manner and to significantly generalize several results in
  black hole physics. It also leads to a number of applications in
  mathematical general relativity, numerical relativity, astrophysics,
  and quantum gravity. In this short review, I will discuss the basic
  ideas and recent developments in this framework, and summarize some
  of its applications with an emphasis on numerical relativity.
\end{abstract}

\section{Introduction}
\label{sec:intro}

The surface of a black hole has traditionally been defined using event
horizons.  Event horizons play a fundamental role in many seminal
investigations in black hole physics.  This includes Hawking's area
increase theorem, black hole thermodynamics, the uniqueness theorems,
black hole perturbation theory and the topological censorship results.
Moreover, the most important family of black holes for many purposes
are the Kerr-Newman black holes.  Similarly for almost all
astrophysical purposes, most studies are carried our using Kerr black
holes.  Given this list of successful results and applications, is
there any real need to go beyond event horizons and Kerr black holes?

There are indeed some situations where event horizons are not
sufficient, and most of these have to do with the global nature of
event horizons; we need to know the entire history of the spacetime in
order to locate them.  This leads to a practical problem for numerical
relativity simulations.  There is no way to locate event horizons
using only Cauchy data at a given time without actually performing the
simulation and constructing the full spacetime.  Moreover, even after
the event horizon is located, using it to calculate the physical
parameters is fraught with difficulties.  In particular, the
Hamiltonian methods used to define the black hole parameters as
generators of symmetries are not well adapted to the event horizon.
All these problems are resolved in the case when the spacetime is
stationary.  However we would like to go beyond stationarity, and even
for black holes in equilibrium, it should not be necessary to require
the entire spacetime to be stationary.

One of the classic results of crucial importance to black holes which
does not use event horizons are the singularity theorems of Penrose
and Hawking \cite{Penrose:1964wq,Hawking:1969sw}. The presence of a
closed trapped surface implies geodesic incompleteness in the future.
The first singularity theorem was proved by Penrose in 1965
\cite{Penrose:1964wq}, and this paper also introduced the notion of a
trapped surface.  We shall use Penrose's trapped surfaces to study
black holes quasi-locally, without relying on global properties of the
spacetime.

The rest of this review is organized as follows.  Following a
discussion of basic notions and definitions, we discuss the existence
and non-uniqueness of quasi-local horizons, and the time evolution of
marginally trapped surfaces in Sec.~\ref{sec:properties}.
Sec.~\ref{sec:secondlaw} discusses the black hole area increase law.
Finally Sec.~\ref{sec:numrel} describes some applications in numerical
relativity.  The reader should beware that this is a biased review of
quasi-local horizons with a focus on numerical relativity
applications.  There are a number of other interesting mathematical
and physical aspects of quasi-local horizons which we shall not have
time to discuss.  The reader is referred to
\cite{Ashtekar:2004cn,Gourgoulhon:2005ng,Booth:2005qc} for more
complete reviews and references.

\subsection*{Trapped surfaces and the trapping region}

The expansion of a congruence of null geodesics is defined as the rate
of increase of an infinitesimal transverse 2-dimensional cross-section
area $\delta A$ carried along with the geodesics:
\begin{equation}
  \label{eq:1}
  \Theta = \frac{1}{\delta A}\frac{d \delta A}{dt}\,.
\end{equation}
The definitions of the shear $\sigma_{ab}$ and twist $\omega_{ab}$ are
also based on deformations of the cross-section.  The particular
geodesic congruence we consider are the ones orthogonal to a 2-surface
$S$.  Let us denote the in-going and out-going null normals to $S$ by
$\ell^a$ and $n^a$ respectively, and let $\Theta_{(\ell)}$ and
$\Theta_{(n)}$ be their respective expansions.  For a sphere in flat
space, the out-going light rays are diverging and the ingoing ones are
converging, i.e. $\Theta_{(\ell)} >0$ and $\Theta_{(n)}<0$.  $S$ is
said to be a trapped surface if both sets of null-normals are
converging: $\Theta_{(\ell)}<0$ and $\Theta_{(n)}<0$.  A marginally
trapped surface (MTS) is one for which $\Theta_{(\ell)} = 0$ and
$\Theta_{(n)}<0$.  As shown by the singularity theorems, the presence
of such surfaces is the signature of a spacetime containing a black
hole.  Note however that this is not necessarily a signature of strong
gravitational field; they are present even for large black holes which
have correspondingly small tidal forces at the horizon.  It can be
shown that trapped surfaces must lie inside the event horizon, and
that cross-sections of the event horizon for stationary black holes
are MTSs.  

The spacetime region $\mathcal{T}$ containing trapped surfaces is called the
trapped region.  Similarly, if we restrict our attention to a
initial-data surface $\Sigma$, and to trapped surfaces lying on
$\Sigma$, we can similarly define the trapped region $\mathcal{T}_\Sigma
\subset \Sigma$ which is, by definition, a subset of the full
four-dimensional trapped region.  An apparent horizon is the outermost
MTS on $\Sigma$.  It can be shown that the MTSs form the boundary of
the trapped region on $\Sigma$.  It seems plausible that the event
horizon should be the boundary of $\mathcal{T}$.  This is indeed the case in
Schwarzschild where there exist spherically symmetric trapped surfaces
all the way up to the event horizon, and the cross-section of the
event horizon is a MTS.  Thus, the event horizon separates the trapped
and normal region of spacetime. This is also true for the Kerr black
hole.  However, in dynamical spacetimes, the event horizon is growing,
and thus cross-sections of the event horizon have $\Theta_{\ell)}> 0$.
Thus, there cannot be a sequence of trapped surfaces which approach
smoothly to the event horizon.  Is it then still true that the event
horizon is the boundary of the trapped region?  It was proposed by
Eardley that it is indeed the case \cite{de}, and that trapped
surfaces can be deformed to get arbitrarily close to the event horizon
(but the limit itself is not smooth).  A useful toy example in
spherical symmetry is the Vaidya spacetime which models the collapse
of null dust.  In this case, it is seen that the spherically symmetric
trapped surfaces do not extend all the way to the event horizon.
However, Eardley's conjecture, if true, implies that there exist
non-spherically symmetric trapped surfaces which extend up to the
event horizons.  Numerical evidence was provided in
\cite{Schnetter:2005ea} and later proved analytically by Ben-Dov that
this is indeed the case \cite{BenDov:2006vw}.
\begin{figure}
  \psfrag{scrip}{\tiny{$\mathcal{I}^+$}}
  \psfrag{scrim}{\tiny{$\mathcal{I}^-$}}
  \psfrag{ip}{\tiny{$i^+$}}
  \psfrag{im}{\tiny{$i^-$}}
  \psfrag{i0}{\tiny{$i^0$}}
  \psfrag{eh}{\tiny{$E$}}
  \psfrag{v0}{\tiny{$v=0$}}
  \psfrag{r0}{\tiny{$r=0$}}
  \psfrag{dh}{\tiny{$H$}}
  \psfrag{ev}{$E$}
  \psfrag{tr}{$\,\,S$}
  \includegraphics[width=0.4\textwidth]{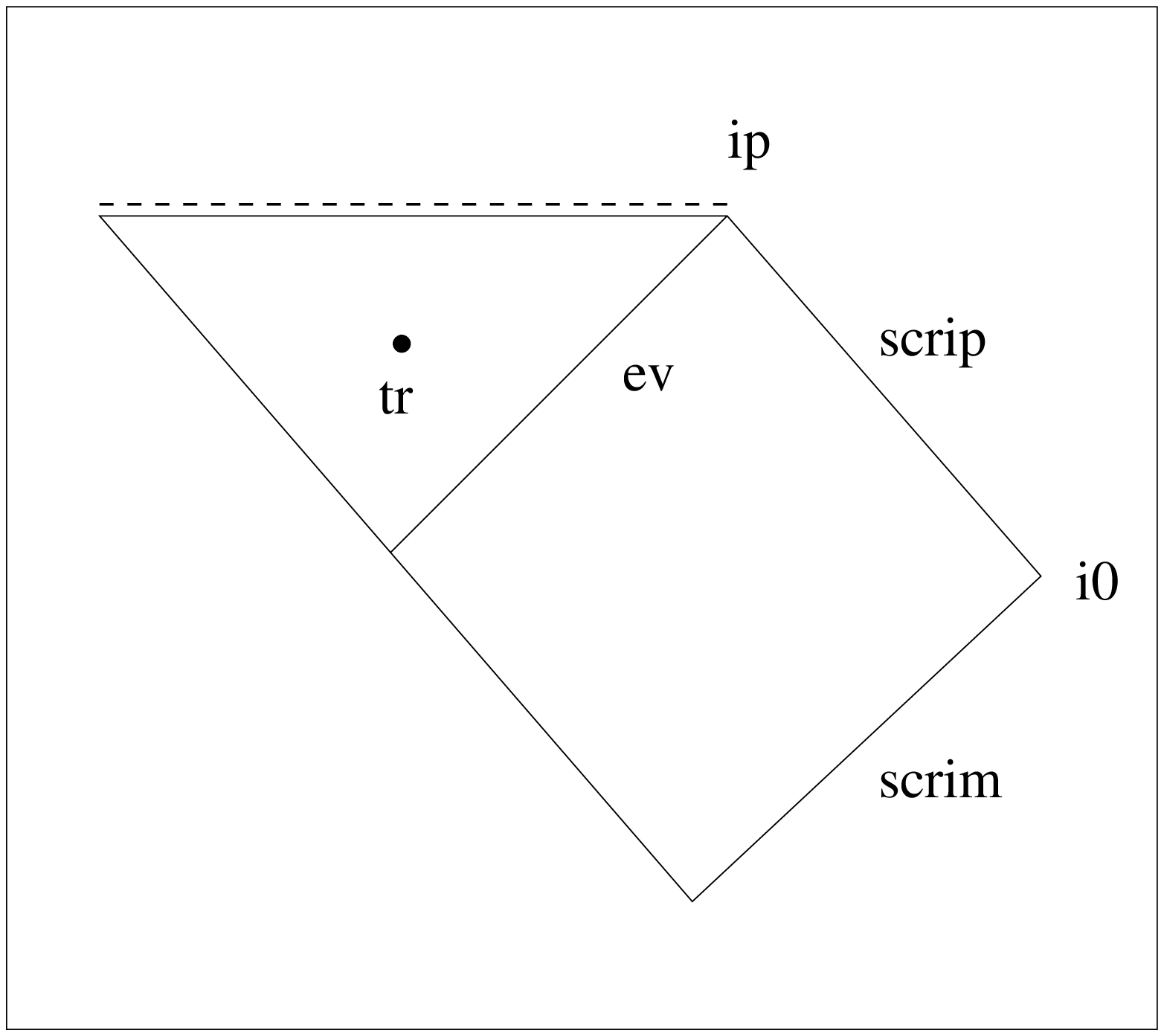}
  \includegraphics[width=0.3\textwidth]{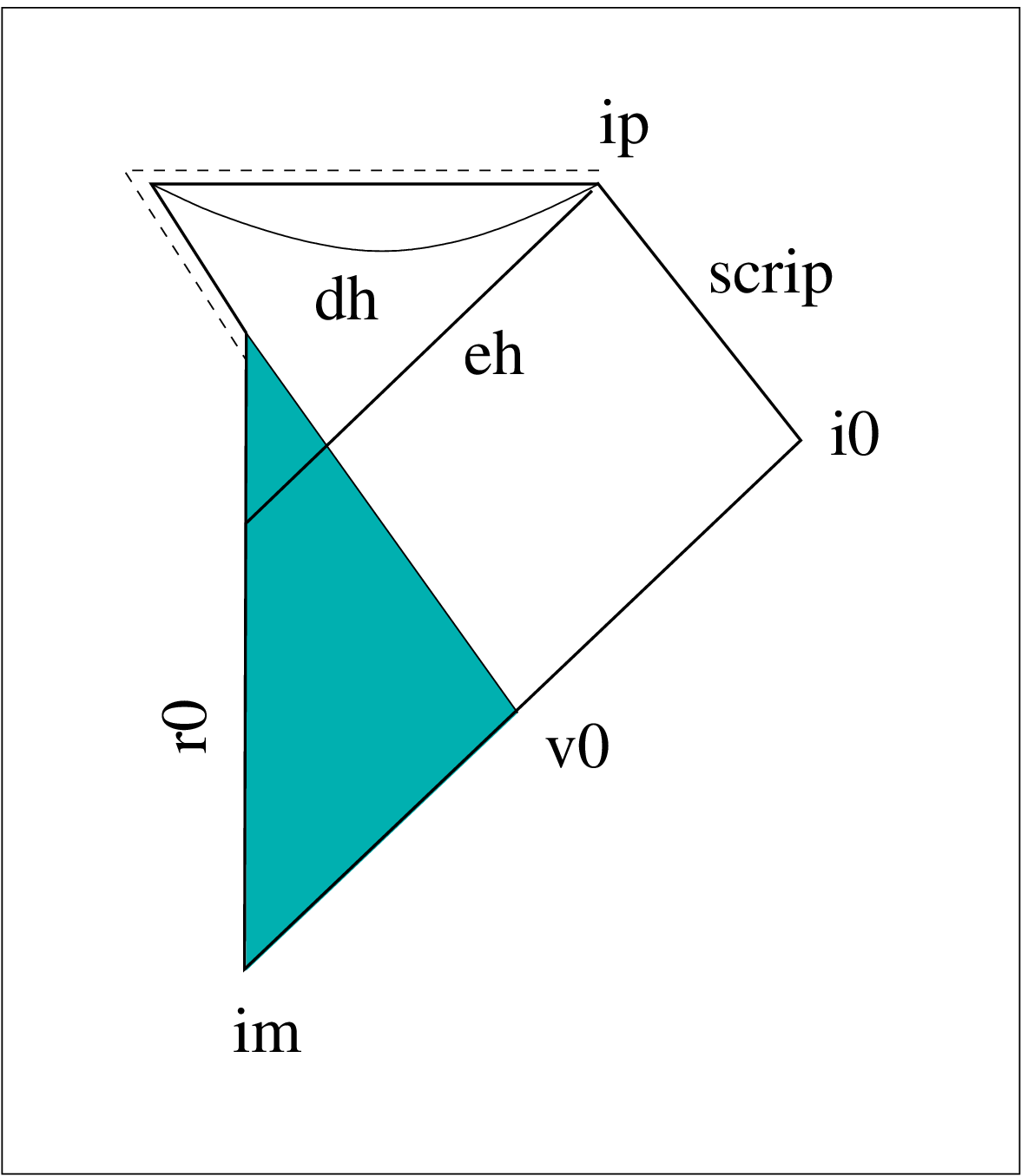}
  \caption{Conformal diagrams for (half of) Schwarzschild (left panel)
    and for the Vaidya spacetime (right panel). In Schwarzschild,
    there exist spherically symmetric MTSs all the way up to the event
    horizon. The shaded region in the right panel is flat Minkowski
    space. The collapse begins at a retarded time $v=0$ and forms a
    black hole.  The analog of the Schwarzschild $r=2M$ surface is now
    spacelike and is denoted by $H$. Spherically symmetric MTSs cannot
    cross $H$.}
\end{figure}
Trapped surfaces form the basis of recent developments in the study of
quasi-local horizons.  One of the first papers which took trapped
surfaces seriously was by Sean Hayward in 1994 who introduced the
notion of a trapping horizon \cite{sh} which will be discussed later.
Black holes in equilibrium were studied by Ashtekar, Beetle and
Fairhurst \cite{abf1,abf2,ihprl} with the aim of using them in quantum
gravity black hole entropy calculations \cite{abk}.  They have been
subsequently developed further and have been useful in other areas
such as black hole mechanics
\cite{abl2,afk,booth,Booth:2005ss,Booth:2001gx,adw,Korzynski:2004gr,Liko:2007th},
numerical relativity, the study of hairy black holes (see e.g.
\cite{acdilaton,acs,acs2,kksw,kk1}) etc.

\section{Fundamental properties of black hole horizons}
\label{sec:properties}

Let us outline some basic definitions and properties of quasi-local
horizons. The starting point for most of these constructions is the
notion of a marginally trapped tube (MTT) defined to be a
three-surface of topology $S^2\times \mathbb{R}$ foliated by MTSs. It
is useful to think of a MTT as being obtained by the time evolution of
a MTS.  The MTT is thus constructed by stacking up MTSs found at
different times.  The various kinds of quasi-local horizons are MTTs
with additional conditions on whether it is a spacelike, timelike or
null surface, and additional geometric requirements on $\Theta_{(n)}$.
\begin{center}
  \begin{tabular}[t]{l|l|l}
    Name & Signature & Additional conditions on $\Theta_{(n)}$ and $\Theta_{(\ell)}$\\
    \hline
    Isolated horizon & Null & \parbox{3in}{None (but with additional
    conditions on other geometrical fields)}\\
    Dynamical horizon & Spacelike & $\Theta_{(n)} < 0$ \\
    Timelike membrane & Timelike  & None\\
    \parbox{1.5in}{Trapping horizon\\ (future outer)} & No restriction & $\Theta_{(n)} < 0$ and
    $\mathcal{L}_n\Theta_{(\ell)} < 0$
  \end{tabular}
\end{center}
These various constructions are relevant in different circumstances.
An isolated horizon models an isolated black hole in an otherwise
dynamical spacetime.  The event horizon of stationary black holes are
isolated horizons.  In a dynamical case, dynamical horizons or
future-outer-trapping-horizons are the most relevant.  Timelike
membranes, as their name indicates, are timelike surfaces. Thus, they
are not one-way membranes and do not function as valid black hole
surfaces. They do however exist and can be found, for example, in
numerical simulations of black hole mergers.  Useful examples of these
different types of horizons can be found in, e.g.
\cite{Booth:2005ng,Schnetter:2006yt,Nielsen:2005af}.

The key applications of these notions have been in the black hole
entropy calculations, in formulating a more general framework for
black hole mechanics, in studying properties of hairy black holes with
non-Abelian gauge fields or dilatons, and in helping to study
mathematical properties of trapped surfaces.  They have also been
found to be useful in astrophysical context primarily through
numerical simulations of black hole spacetimes.

\subsection*{The existence and non-uniqueness of quasi-local horizons}

Let us now briefly discuss the time evolution of MTSs.  Numerically,
it is sometimes seen that apparent horizons behave discontinuously,
and this is perhaps the main reason why they have not been taken
seriously in the past.  However, it turns out that the behavior of
MTSs themselves is smooth; there is, to my knowledge, no known example
when this is not the case.  The discontinuous time evolution of
apparent horizons is an artifact of the ``outermost'' condition
appearing in the definition of an apparent horizon.  There is now in
fact a rigorous mathematical proof by Andersson, Mars and Simon that
MTSs subject to a stability requirement (which is expected to hold for
the physically most important cases) evolve smoothly
\cite{Andersson:2007fh,Andersson:2005gq} (see also
\cite{Booth:2006bn}).  We briefly sketch the statement of the result.

An MTS $S$ on a spatial slice $\Sigma$ is said to be
strictly-stably-outermost if there exists an infinitesimal first order
outward deformation which makes $S$ strictly untrapped.  Explicitly,
if $\mathbf{r}$ is the unit spacelike normal to $S$ on $\Sigma$, then
we consider displacements of $S$ (and geometric fields on $S$) along
$f\mathbf{r}$ for some function $f$; outward deformations have $f\geq
0$.  Then $S$ is strictly-stably-outermost if the first order
variation of the expansion $\Theta_{(\ell)}$ is positive:
$\delta_{f\mathbf{r}}\Theta_{(\ell)} > 0$ for $f\geq 0$.  A crucial
tool in these results is the stability operator $L_{\mathbf{r}}[f] :=
\delta_{f\mathbf{r}}\Theta_{(\ell)}$ which turns out to be an elliptic
operator, and the stability condition can be recast as a condition on
the principal eigenvalue of $L$; see also If this stability condition
is satisfied, then the MTT produced by the time evolution of $S$
exists at least for a sufficiently short duration, and it continues to
exist as long as this stability condition holds. Furthermore, the MTT
in the neighborhood of $S$ is either null or spacelike.  It is
spacelike of the matter flux $T_{ab}\ell^a\ell^b$ is non-vanishing
somewhere on $S$.  The elliptic nature of $L$ ensures that the MTT is
spacelike everywhere in a neighborhood of $S$ if the flux
$T_{ab}\ell^a\ell^b$ is non-zero even in a very small region on $S$.
It should be emphasized that these results do not imply that the time
development of $S$ is unique.  It in fact implies quite the opposite:
for every choice of time evolution by a foliation by spacelike
surfaces (i.e. for every choice of lapse and shift functions) there
exists an MTT and the different MTTs constructed from the different
gauge choices are, in general, distinct from each other.  It is also
worth noting that not all MTSs will satisfy the stability condition;
the unstable MTSs will be inner horizons and actually occur quite
frequently in numerical simulations \cite{Schnetter:2006yt} (though
they are usually not looked for).  However, even the unstable MTSs
always seem to evolve smoothly as far as the numerical simulations are
concerned.  This leads us to believe that the existence result might
be of more general validity and it might be possible to extend the
above techniques to prove this.  See
\cite{Andersson:2005me,Metzger:2007ik,Andersson:2007gy} for further
results on trapped surfaces and quasi-local horizons using similar
techniques.  See also \cite{Szilagyi:2006qy} for interesting numerical
results on the behavior of MTTs in binary black hole spacetimes.  

Complementary to these existence results, there are other important
results on dynamical horizons worth mentioning.  In
\cite{Ashtekar:2005ez} it is proved that the foliation of a dynamical
horizon $H$ by MTSs is unique.  This, together with the existence
results above implies that for a given MTSs on an initial slice
$\Sigma$, the MTTs corresponding different time developments must
really be distinct as 3-manifolds.  There are also some restrictions
on the location of the various dynamical horizons.  For example, it is
shown in \cite{Ashtekar:2005ez} that for a given dynamical horizon
$H$, there cannot be any closed MTSs (and thus no other DH) lying in
the past domain of dependence of $H$.  Thus, while DHs are far from
unique, there are some restrictions on where they can occur.  We
briefly mention results regarding the (non-)existence of dynamical
horizons in spacetimes with symmetries
\cite{Ashtekar:2005ez,Mars:2003ud,Carrasco:2007tn}.  For example, it
is shown in \cite{Mars:2003ud} that strictly stationary spacetime
regions cannot contain trapped or marginally trapped surfaces, and
thus no quasi-local horizons as well.  Finally, a different approach
to studying dyamical horizons is presented in \cite{Bartnik:2005qj}
which considers the conditions on the Cauchy data on $H$ that must be
satisfied if $H$ is a dynamical horizon.  This can be studied in
spherical symmetry, and it leads to necessary conditions for the
spacetime to contain a dynamical horizon; in this regard, see also
\cite{Williams:2007tp}.

\section{The second law}
\label{sec:secondlaw}

In this section, we outline the second law for dynamical/trapping
horizons and its ramifications.  But before doing so, it is worth
mentioning the significant amount of work devoted to understanding the
first law for quasi-local horizons.  The fist law connects variations
in the mass $M$ between two nearby black hole solutions to the surface
gravity $\kappa$, area $A$, angular velocity $\Omega$ and angular
momentum $J$ (the presence of other conserved charges is easy to
incorporate): 
\begin{equation}
  \label{eq:2}
  \delta M = \frac{\kappa}{8\pi G}\delta A + \Omega\delta J\,.
\end{equation}
This was initially proved for stationary Kerr black holes. It has been
generalized to isolated horizons \cite{afk}. This is a significant generalization
and it leads to a better understanding of the nature of the first law.
It has since also been further extended to include a physical process
version for dynamical horizons.  There is also an earlier treatment by
Hayward based on a different formulation.

Let us now turn to the second law.  It was originally formulated by
Hawking as: if matter satisfies the null energy condition, then the
area of an event horizon can never decrease, i.e. $\Delta A \geq 0$.
This is an exact result in full general relativity without any
approximations. It suggests the identification of black hole area with
entropy, and this identification has been very important as a driving
force for progress in quantum gravity.

Classically, the second law leads to the picture of a black hole
growing inexorably as it swallows matter and radiation.  We can thus
ask whether it is possible to get an equation like 
\begin{equation}
  \label{eq:3}
  \Delta A = \textrm{flux of matter + radiation}\,.
\end{equation}
We require the fluxed to be quasi-local, geometric, and positive
definite.  The approach of Hartle and Hawking using perturbation
theory and event horizons is a significant step, and it reinforces
the possibility of the general validity of such a formula.  It is easy
to see however that if we restrict ourselves to event horizons, it is
impossible to obtain in full generality because of, again, the
global nature of the event horizon. A simple example is the Vaidya
spacetime; in the shaded region before the null-radiation has formed
the black hole, the event horizon is growing in anticipation of the
black hole forming in the future.  Thus, the event horizon is growing
in flat space when surely any reasonable definition of the fluxes must
be zero.  

This non-locality can be seen in a different way.  We can ask whether
it is possible to obtain a differential equation for the rate of area
increase.  Working in the membrane paradigm, for event horizons, it
was shown by Damour that
\begin{equation}
  \label{eq:4}
  \frac{d^2A}{dt^2} - \kappa \frac{dA}{dt} = \ldots \qquad
  \textrm{with}\qquad \kappa > 0\,.
\end{equation}
The right hand side of this equation contains the source terms from
the infalling matter/radiation which causes the black hole to grow,
and all terms in thus equation are quasi-local. The sign of $\kappa$
is however a problem; it will generically lead to an exponential
divergence if we attempt to solve (\ref{eq:4}) as an initial value
problem.  In fact, we need to impose $dA/dt \rightarrow 0 $ as
$t\rightarrow \infty$ to get finite solutions.  

Can we reformulate the second law for quasi-local horizons?  In fact,
$\Delta A > 0$ is a simple consequence of $\Theta_{(\ell)}=0$ and
$\Theta_{(n)} < 0$ for dynamical horizons.  We can do better.
Performing the decomposition of all geometric fields on a dynamical
horizon, and using the constraint equations on a dynamical horizon, it
can be shown that 
\begin{equation}
  \label{eq:5}
  \frac{R_2}{2G}-\frac{R_1}{2G} = \mathcal{F}^{(R)}_m +
  \mathcal{F}^{(R)}_g 
\end{equation}
where 
\begin{equation}
  \label{eq:6}
  \mathcal{F}^{(R)}_g := \frac{1}{16\pi G}
  \int_{\Delta H} N_R\left\{ |\sigma|^2 + 2|\zeta|^2\right\}\,d^3V  
\end{equation}
is the flux of the infalling gravitational radiation.  The flux
$\mathcal{F}^{(R)}_g$ has a number of reasonable properties. All terms
appearing in the integral are local and coordinate independent, it is
manifestly non-negative and it vanishes in spherical symmetry.

It is also possible to formulate a differential law for the area
increase.  Using a different component of the Einstein equation than
what was used to get (\ref{eq:5}), Gourgoulhon and Jaramillo
\cite{Gourgoulhon:2006uc} obtained an equation of the following form:
\begin{equation}
  \label{eq:7}
    \frac{d^2A}{dt^2} + \kappa^\prime \frac{dA}{dt} = \ldots \qquad
  \textrm{with}\qquad \kappa^\prime > 0\,.
\end{equation}
This is similar to (\ref{eq:4}), all terms in it are local and again
the right hand side contains the source terms corresponding to
infalling matter/radiation fluxes. However, the $dA/dt$ term now
appears with the opposite sign, and we can therefore indeed solve this
differential as an initial value problem by specifying $A$ and $dA/dt$
at some initial time $t=0$.  The solution at some $t$ depends only on
the fields at earlier times, and there is thus no teleological
problem.  This approach also solves other problems with the membrane
paradigm. For example, the bulk viscosity of the horizon becomes
positive \cite{Gourgoulhon:2005ch,Gourgoulhon:2006uc} as for a general
fluid.  Thus, consistent with the membrane paradigm, a black hole
horizon can indeed be treated as a normal object and we can assign
physical properties to it.

\section{Applications in numerical relativity}
\label{sec:numrel}

Marginally trapped surfaces have been used in numerical simulations of
black hole spacetimes almost from the very beginning of the field.
This is because it is important to keep track of the black hole(s)
while the simulation is in progress.  The global nature of event
horizons makes them not very useful for this purpose because, as
discussed earlier, we need the full spacetime to locate the event
horizon; the spacetime is the end product of the simulation, and is
not available to us in real-time while the simulation is in progress.
Thus, it is clear that quasi-local horizons can be useful in numerical
relativity.  They have so far been primarily used to extract gauge
invariant information about the black hole such as its mass, angular
momentum etc., and this is what we shall mostly focus on in this
section.  However, there are also other important applications which
we will not be able to discuss.  This includes the construction of
initial data with trapped and marginally trapped surfaces as the inner
boundary
\cite{jgm,djk,sd2,Maxwell:2003av,Gourgoulhon:2007tn,Smith:2007gg}, and
the possibility of using fully constrained evolution schemes with a
dynamical horizon as the inner boundary \cite{Jaramillo:2007km}.
There has also been interest in clarifying the definition of surface
gravity pf a quasi-local horizon, and the closely related notion of
extremality \cite{Booth:2007ix,Booth:2007wu,Nielsen:2007ac}.  This
could prove to be useful for mathematical and astrophysical
applications.

\subsection*{Mass and angular momentum}

Numerical simulations are based on the initial value formulation of
general relativity. Thus we are given an initial data set
$(\Sigma,h_{ab},K_{ab})$ where $\Sigma$ is a 3-manifold embedded in
the full spacetime, $h_{ab}$ a Riemannian metric on $\Sigma$, and
$K_{ab}$ the second fundamental form describing how $\Sigma$ is
embedded in the spacetime.  Such an initial data set is evolved in
time to construct the full spacetime.  There are various formalisms
for performing these evolutions and there are different choices of the
variables that can be evolved, but these issues are not of much
concern for our purposes.  We instead pose a straightforward question.
Assuming $\Sigma$ to be some slice of the Kerr spacetime, and assuming
that $\Sigma$ intersects the event horizon in a complete
sphere\footnote{And also assuming that we have an efficient way of
  locating marginally trapped surfaces on $\Sigma$.}, how can we
determine its parameters, i.e. its mass and angular momentum?
Depending on the choice of $\Sigma$ and the choice of coordinates on
$\Sigma$, the shape of the apparent horizon may turn out to be quite
complicated. It may not seem axisymmetric and it may even be difficult
to say whether we have a stationary black hole.

Let us reformulate this question in a much more general context.
Consider a quasi-local horizon $H$ (i.e. an isolated, dynamical or
trapping horizon) and let us assume that $\Sigma$ intersects $H$ in a
cross-section $S$ which is a marginally trapped surface.  We assume
that $\Sigma$ is an asymptotically flat slice with $S$ as its inner
boundary (the generalization to multiple black holes, i.e. when $S$
consists of several disconnected components is straightforward).  Let
us assume that $H$ is axisymmetric, i.e. it has a rotational vector
$\varphi^a$ which has closed orbits, vanishes at two points on each
cross-section of the MTSs which foliate $H$, and which is a symmetry
of the geometrical fields on $H$. In particular $\varphi^a$ is a
symmetry of the two metric $q_{ab}$ on every cross-section $S$ of $H$.
Note that we only asked for $\varphi^a$ to exist on $H$, and not in
the full spacetime and not even in a neighborhood of $H$.  It is then
possible to associate an angular momentum $J^{(\varphi)}_S$ of the
horizon.  Just like in classical mechanics where conserved quantities
are defined as generators of symmetries, this calculation is based on
a Hamiltonian formalism.  We calculate the generator of
diffeomorphisms along a rotational vector field $\phi^a$ which
coincides with $\varphi^a$ on $H$ and with an asymptotic rotational
symmetry at infinity \cite{abl2,afk,booth,Booth:2005ss,Booth:2001gx}
(analogous calculations also work in 2+1 \cite{adw} and higher
\cite{Korzynski:2004gr,Liko:2007th} dimensions).  It is then easy to
identify the contribution of the horizon to the angular momentum, and
it turns out to be given by a surface integral over $S$ \cite{dkss}:
\begin{equation}
  \label{eq:8}
  J_S^{(\varphi)} = \frac{1}{8\pi} \oint_{S} K_{ab}\varphi^a dS^b  \,.
\end{equation}
Note that this formula is analogous to the standard formula for the
angular momentum at spatial infinity (which is in fact also found by
the same Hamiltonian calculation).

For an isolated horizon, as expected, $J^{(\varphi)}_S$ turns out to
be independent of $S$.  In this case, it is also easy to show that
\emph{every} cross-section of $H$ is actually
axisymmetric\footnote{This may seem surprising because this is
  certainly not the case for a normal $S^2\times\mathbb{R}$ cylinder
  in Euclidean space.  If there is a symmetry vector $\varphi^a$ on
  the cylinder, it need not project to a symmetry vector on a given
  cross-section $S$ of the cylinder.  It is nevertheless true for an
  isolated horizon because if $\varphi^a$ is a symmetry, then so is
  $\varphi^a + f\ell^a$ for any function $f$ and null generator
  $\ell^a$; projecting $\varphi^a$ to $S$ is equivalent to a
  particular choice of $f$.}  It is now clear how the angular momentum
associated with $S$ should be calculated.  Having located $S$ on
$\Sigma$, we need to find the appropriate symmetry vector on $S$ and
then calculate the surface integral (\ref{eq:8}).  There are now a
number of methods proposed for calculating $\varphi^a$
\cite{dkss,Cook:2007wr,Korzynski:2007hu}.   

With this background, the answer to the question posed at the
beginning of this section is clear.  If $S$ is a cross-section of the
Kerr horizon, then it is guaranteed to have an axial symmetry vector
$\varphi^a$ which can then be used to calculate the angular momentum
via (\ref{eq:8}). Then given the angular momentum, we can calculate
the area of the horizon thereby identifying the Kerr solution
uniquely.  

Given the angular momentum and the horizon area $A_S$ (and the area
radius $R_S = \sqrt{A_S/4\pi}$, the horizon mass $M_S^{(\varphi)}$ is
\begin{equation}
  \label{eq:9}
  M_S^{(\varphi)} = \frac{1}{2R_S}\sqrt{R_S^4 + 4(J_S^{(\varphi)})^2} \,.
\end{equation}
This will give the correct answer for Kerr, and it is in fact also the
result of a Hamiltonian calculation in the more general case of an
axisymmetric quasi-local horizon.  

Apart from getting the magnitude of the spin, it is also possible to
estimate the direction of the angular momentum vector. The basic idea
is to use the poles of $\varphi^a$ to define the axis of rotation.
While the poles themselves are well defined on a given quasi-local
horizon, the procedure of assigning a vector is not as clear cut.  For
example, it is not clear how the spin direction thus obtained can be
compared with the spin direction calculated at spatial infinity.
Nevertheless, this method has been applied and preliminary results
are promising \cite{Campanelli:2006fy}.  

Equation (\ref{eq:8}) is now being used fairly widely in numerical
relativity, though there is possibly room for improvement in the
calculation of the symmetry vector field $\varphi^a$ along the lines
of \cite{Cook:2007wr}, for having a better conceptual understanding of
the meaning of $J_S^{(\varphi)}$ when $\varphi^a$ is only an
approximate symmetry vector (which is invariably the case in numerical
simulations), and also a better understanding of the spin direction
which is important for astrophysical applications.

\subsection*{Higher multipole moments}

Apart from calculating the angular momentum and mass, it turns out
that it is also possible to meaningfully define the higher multipole
moments of a quasi-local horizon, at least in the axi-symmetric case.
This construction was first carried out by \cite{aepv} for isolated
horizons, and subsequently applied by
\cite{Ashtekar:2004cn,Schnetter:2006yt} to the general case of a
dynamical horizon.  The construction starts with a coordinate system
built using the given axial vector $\varphi^a$.  $\phi\in [0,2\pi)$ is
affine parameter along $\varphi^a$ and $\zeta = \cos\theta$ defined by
$d\zeta \propto \star\varphi$ (the proportionality factor is chosen by
requiring $\oint_S\zeta d^2V = 0$). Then we use the spherical
harmonics in these $(\theta,\phi)$ coordinates to define the mass and
current multipole moments:
\begin{equation}
  \label{eq:11}
  M_n^{(\varphi)} = \frac{R_S^n
    M_S^{(\varphi)}}{8\pi}\oint_S\left\{\tilde{{R}}P_n(\zeta)\right\} d^2V\,, \quad
  J_n^{(\varphi)} = \frac{R_S^{n-1}}{8\pi} \oint_S
  P_n^\prime(\zeta)\bar{K}_{ab}\varphi^ad^2S^b
\end{equation}
where $P_n$ is the $n^{\rm th}$ order Legendre polynomial and
$P_n^\prime$ its derivative.  

These equations provides the source multipole moments of black hole,
which are in general distinct from the field multipole moments defined
at infinity.  For isolated horizons, it can be shown \cite{aepv} that
the intrinsic horizon geometry is completely characterized by these
multipole moments, i.e. any two isolated horizons with the same
multipole moments are diffeomorphic to each other.  $J_0$ vanishes by
absence of monopole (NUT) charges, $M_0$ is mass and $J_1$ is angular
momentum.  In Kerr, $M_0$ and $J_1$ determine all higher moments.  In
Schwarzschild, only $M_0 \neq 0$.  The higher moments provide a
convenient way of quantifying the deviation from Kerr. These multipole
moments were applied in some example numerical simulations in
\cite{Schnetter:2006yt} where it was shown that the black holes do
indeed converge to Kerr very quickly after merger. However, the
simulations were unfortunately not accurate enough to extract the late
time decay rates of the multipole moments which would be the analog of
Price's law for dynamical horizons.  Hopefully this can be measured in
the future using long duration accurate simulations.

\subsection*{Quasi-local linear momentum}

The calculation of black hole linear momentum is of astrophysical
importance in the context of the recoil velocity produced during the
merger of two black holes.  The reason for the recoil is the
anisotropic emission of gravitational radiation.  It has been found
that certain initial spin configurations lead to a much larger than
expected value of this recoil velocity and the largest contribution
turns out to be from the merger phase.  This result is especially
interesting for the case of super-massive black holes; if the recoil
is large enough, the remnant black hole may be kicked out of the host
galaxy and this has important astrophysical implications.  Most
calculations of the recoil velocity are based on the gravitational
waveform extracted far away from the black hole which measures the
center-of-mass momentum of the system.  It is thus natural to ask
whether one can measure the momentum quasi-locally for the two
individual black holes.  This would be a useful consistency check and
it could also give us more detailed information about the dynamics of
the merger.  The possibility of measuring the linear momentum, and
more generally the quasi-local energy-momentum four-vector, is also of
mathematical interest.

So far, we have justified the equations for angular momentum, mass and
energy by Hamiltonian methods.  For the angular momentum we assumed
the existence of a rotational symmetry and for energy and mass we need
to pick out a preferred time evolution vector field at the horizon.
Following the same line of reasoning, one might think of defining
linear momentum by assuming the existence of a translational symmetry
in a neighborhood of the horizon, or at least some preferred
translational vector field.  While it might be possible to do this in
special cases, for example when the data is conformally flat, it is
clearly not something we can assume generally.  Unlike for angular
momentum where there are interesting regimes where approximate
axisymmetry is a valid assumption, the basis for carrying over this
approach to linear momentum is much less secure.

Let us then try a more heuristic approach.  Just as the formula for
horizon angular momentum is analogous to the angular momentum at
spatial infinity, let us apply the formula for linear momentum at
infinity to the horizon.  At spatial infinity, for a given asymptotic
translational Killing vector field $\xi^a$, the momentum is
\begin{equation}
  \label{eq:10}
  P_\xi^{\rm ADM} = \frac{1}{8\pi}\oint_{S_\infty} (K_{ab} -
  Kh_{ab})\xi^a dS^b
\end{equation}
where $S^\infty$ is the sphere at infinity.  This motivates the
following definition at the horizon:
\begin{equation}
  \label{eq:12}
  P_\xi^{(S)} = \frac{1}{8\pi}\oint_{S} (K_{ab} -
  Kh_{ab})\xi^a dS^b
\end{equation}
where $\xi^a$ is some translational vector at the horizon.  Using the
constraint equations of $\Sigma$, it is easy to show
\begin{equation}
  \label{eq:13}
  P_\xi^{\rm ADM} - P_\xi^{(S)} = \frac{1}{16\pi}\int (K^{ab} -
  K\gamma^{ab})\mathcal{L}_\xi h_{ab} d^3V\,.
\end{equation}
The right hand side vanishes in some special cases, e.g. when $\xi^a$
is a conformal Killing vector on $\Sigma$, and the data is maximal
($K=0$). We also take $\xi^a$ to be coordinate basis vectors to get
the three components of linear momentum.  This definition is clearly
gauge dependent, but it does seem sufficiently worthwhile to try it
out in a numerical simulation.  This was done in
\cite{Krishnan:2007pu}, and the preliminary results are encouraging.

\begin{figure}
  \centering
  \includegraphics[width=0.45\textwidth]{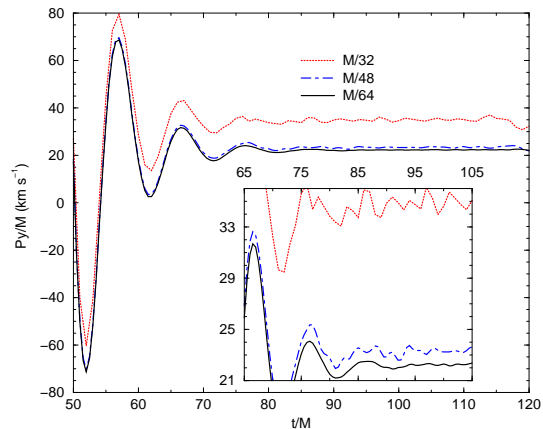}  
  \caption{Linear momentum of the remnant black hole produced by the
    headon merger of black holes with anti-aligned spins for different
    resolutions.}
  \label{fig:linmom}
\end{figure}

An example result is shown in fig.~\ref{fig:linmom} for a head on
collision of two black holes with equal masses, and spins orthogonal
to the line joining the two black holes and oppositely aligned with
magnitude $0.15M^2$.  The remnant black hole is non-spinning and the
expected recoil velocity is $20.4\,$km/s.  Figure \ref{fig:linmom}
shows the result of applying (\ref{eq:12}) at different resolutions.
The extrapolated result turns out to be $21.5\,$km/s which is in fairly
good agreement with the calculation at infinity.

\section{Conclusions}
\label{sec:conc}

In this talk we have outlined some properties of quasi-local horizons
with applications to numerical relativity.  We have seen that
marginally trapped surfaces are not as badly behaved as one might have
thought based on intuition about apparent horizons.  It is possible to
prove useful and interesting mathematical results about them and they
can be used to study black hole physics.  They have applications in
diverse fields ranging from quantum gravity to numerical relativity.
In numerical relativity we have shown how one can extract physical
parameters of black holes such as its mass, angular momentum and the
higher source multipole moments as well. We have also discussed
preliminary ideas about the quasi-local linear momentum of black holes
which could be of some astrophysical importance.  On a more general
note, we have seen that numerical relativity can be used as a tool for
proposing and testing mathematical conjectures regarding trapped
surfaces and black holes and in fact more generally problems in
geometric analysis which are difficult to deal with analytically.  

\section*{Acknowledgements}

I am grateful to Abhay Ashtekar for valuable discussions and to the
organizers of GR18 for their hospitality.  

\section*{Bibliography}
\bibliographystyle{unsrt}
\bibliography{refs}

\end{document}